\documentclass[iop,revtex4]{emulateapj}
\usepackage{lineno}

\begin{document}

\renewcommand{\topfraction}{1.0}
\renewcommand{\bottomfraction}{1.0}
\renewcommand{\textfraction}{0.0}

\newcommand{\kms}{km~s$^{-1}$\,}
\newcommand{\msun}{$M_\odot$\,}

\title{Spectroscopic orbits of subsystems in  multiple
  stars. V.}

\author{Andrei Tokovinin}
\affil{Cerro Tololo Inter-American Observatory, Casilla 603, La Serena, Chile}
\email{atokovinin@ctio.noao.edu}

\begin{abstract}
Spectroscopic  orbits  are determined  for  inner  subsystems in  nine
stellar  hierachies  (HIP  2863,  4974, 8353,  28796, 35261,  92929, 
115272,  115552, and  117596). Their  periods  range from  2.5 to  312
days. For  each system, estimates  of masses, orbital  inclination and
other parameters are given.  
\end{abstract}

\maketitle

\section{Introduction}
\label{sec:intro}

Spectroscopic orbits of nine  close binaries belonging to hierarchical
stellar  systems are determined  here,  continuing the  series of
similar  papers   \citep{paper1,paper2,paper3,paper4}.   Knowledge  of
orbital  periods  and other  characteristics  of hierarchical  systems
paves the  way to  constrain and refine  theories of  their formation,
still in  their infancy. Binaries with  periods on the order  of a few
days, studied  here, are  affected by tides,  and the  distribution of
their periods is expected to  reflect the role of tidal interaction in
the formation of  inner subsystems. The current collection  of data on
stellar hierarchies  \citep{MSC} contains many  spectroscopic binaries
with yet unknown orbits. Ongoing work  slowly fills this gap.

\begin{deluxetable*}{c c rr   l cc rr r c }
\tabletypesize{\scriptsize}     
\tablecaption{Basic parameters of observed multiple systems
\label{tab:objects} }  
\tablewidth{0pt}                                   
\tablehead{                                                                     
\colhead{WDS} & 
\colhead{Comp.} &
\colhead{HIP} & 
\colhead{HD} & 
\colhead{Spectral} & 
\colhead{$V$} & 
\colhead{$V-K$} & 
\colhead{$\mu^*_\alpha$} & 
\colhead{$\mu_\delta$} & 
\colhead{RV} & 
\colhead{$\overline{\omega}$\tablenotemark{a}} \\
\colhead{(J2000)} & 
 & &   &  
\colhead{type} & 
\colhead{(mag)} &
\colhead{(mag)} &
\multicolumn{2}{c}{ (mas yr$^{-1}$)} &
\colhead{(km s$^{-1}$)} &
\colhead{(mas)} 
}
\startdata
00363$-$3818  & AB & 2863  & 3365  & G2/G3V  & 8.36 & 1.54 & 247  &  $-$35 &  $-$7.07 & 9.56 \\
01037$-$3024  & AB & 4974  & 6307  & F5/6V   & 8.78 & 1.20 & 35   & $-$34  & $-$2.04   & 13.87\tablenotemark{b} \\
01477$-$4358  & AB & 8353  & 11057 & F6/7V  & 7.97 & 1.34 & 52   & $-$7   & 21.04     & 5.3\tablenotemark{b} \\
06048$-$4828  & A  & 28796 & 41824 & G5V     & 6.57 & 1.75 & $-$101 & $-$23 & 21.35    & 32.51 \\
              & B  & \ldots& \ldots& \ldots  & 7.69 &\ldots& $-$117 & $-$39 & 14.45    & 32.47 \\
07171$-$1202   & AB & 35261 & 56593 & F4V    & 6.71  & 1.14 & 24   & $-$55  & $-$9.0  & 21.6\tablenotemark{b} \\ 
07171$-$1202   & C  & \ldots & \ldots &\ldots& 9.44  & 2.04 & 42   & $-$36  & $-$6.49 & 19.03 \\
18560$-$2503  & A  & 92929 & 175345& G0V     & 7.38 & 1.42 &  38    &51    & 0.16     & 24.20 \\
              & B  &  \ldots& \ldots& \ldots & 12.92&\ldots& 32     & 48    & \ldots   & 22.97 \\
23208$-$5018  & AB & 115272 & 220003& F5m    & 6.05 & 0.94 & 45     & $-$72 & 13.55     & 13.36 \\
              & C  & 115269 & 220002 & G4IV  & 8.86 & 1.40 & 43     & $-$73 & 14.30   & 13.28 \\
23244+1429    & AB & 115552 & 220541 & F8V   & 7.60 & 1.30 & 17     & $-$17 & 16.73   & 6.14 \\
23509$-$7954  & A  & 117596 & 223537 & G3V   & 8.02 & 1.56 & $-$84  & 5     & 23.51   & 18.22 \\
              & B  &  \ldots& \ldots& \ldots & 15.55& 4.70 & $-$86  & 8     & \ldots & 18.11 
\enddata
\tablenotetext{a}{Proper motions and parallaxes are taken
  from the {\it Gaia} DR2 \citep{Gaia}, where available.}
\tablenotetext{b}{{\it Hipparcos} parallax \citep{HIP2}.}
\end{deluxetable*}

The systems  studied here  are listed in  Table~\ref{tab:objects}. The
data are  collected from Simbad  and {\it Gaia} DR2  \citep{Gaia}, the
radial velocities  (RVs) are mostly determined here.  The first column
gives the WDS-style \citep{WDS} code based on the J2000 coordinates
(all objects are actually present in the WDS). 

The structure of this paper is similar to the previous ones.  The data
and methods  are briefly recalled in  Section~\ref{sec:obs}, where the
 orbital elements are also  given. Then each system is discussed in
Section~\ref{sec:obj}.  The  paper  closes  with a  short  summary  in
Section~\ref{sec:sum}.

\section{Observations and data analysis}
\label{sec:obs}

\subsection{Spectroscopic observations}

The spectra used here were taken with the 1.5 m telescope sited at the
Cerro Tololo  Inter-American Observatory (CTIO) in  Chile and operated
by             the             SMARTS            Consortium.\footnote{
  \url{http://www.astro.yale.edu/smarts/}}  The   observing  time  was
allocated  through  NOAO.   Observations  were made  with  the  CHIRON
optical echelle spectrograph \citep{CHIRON} by the telescope operators
in service  mode.  The RVs  are determined from  the cross-correlation
function (CCF)  of echelle  orders with the  binary mask based  on the
solar spectrum, as detailed in \citep{paper1}. The RVs derived by this
method should be  on the absolute scale if  the wavelength calibration
is  accurate.  The  CHIRON RVs  were checked  against standards  and a
small offset of $+0.15$ km~s$^{-1}$  was found in \cite{paper3}; it is
not applied to the RVs given here.

The CCF  contains two dips in  the case of  double-lined systems.  The
dip width is related to the projected rotation speed $V \sin i$, while
its area depends on the spectral type, metallicity, and relative flux.
Table~\ref{tab:dip}  lists average parameters  of the  Gaussian curves
fitted to the CCF dips.   It gives the number of averaged measurements
$N$  (blended  CCFs  were  not  used),  the  dip  amplitude  $a$,  its
dispersion $\sigma$,  the product $a  \sigma$ proportional to  the dip
area (hence to the relative flux), and the projected rotation velocity
$V \sin i$,  estimated from $\sigma$ by the  approximate formula given
in \citep{paper1} and valid for $\sigma < 12$ \kms.  The last column
indicates  the presence  or absence  of the  lithium 6708\AA  ~line in
individual components.

\begin{deluxetable*}{l l c cccc c}    
\tabletypesize{\scriptsize}     
\tablecaption{CCF parameters
\label{tab:dip}          }
\tablewidth{0pt}                                   
\tablehead{                                                                     
\colhead{HIP} & 
\colhead{Comp.} & 
\colhead{$N$} & 
\colhead{$a$} & 
\colhead{$\sigma$} & 
\colhead{$a \sigma$} & 
\colhead{$V \sin i$ } & 
\colhead{Li}
\\
 &  &  & &
\colhead{(km~s$^{-1}$)} &
\colhead{(km~s$^{-1}$)} &
\colhead{(km~s$^{-1}$)} &
\colhead{  6708\AA}
}
\startdata
2863    & B  & 10 & 0.127 & 3.841  &  0.443 &  1.3 & N \\ 
2863    & Aa & 10 & 0.116 & 8.412  &  0.976 & 13.8 & N \\ 
4974    & A  & 11& 0.109 & 4.84  &  0.528 &  6.2 & Y \\
4974    & Ba & 11& 0.080 & 5.79  &  0.465 &  8.4 & N \\
4974    & Bb & 11& 0.030 & 3.67  &  0.111 &  2.5 & N \\
8353    & Aa & 9 & 0.050 & 11.97 &  0.598 & 20.7 & N \\  
8353    & B  & 9 & 0.093 & 5.56  &  0.518 &  7.9 & N \\  
28796   & Aa & 3 & 0.325 & 5.94 &  1.930 & 8.8  & N \\
28796   & B  &  4 & 0.473   & 3.79  & 1.795 & 3.0  & N \\ 
35261   & Ca & 11& 0.418 & 3.66  &  1.53  & 2.4  & N \\ 
35261   & Cb & 11& 0.034 & 5.39  &  0.184 & 7.5:  & N \\ 
92929   & Aa & 5 & 0.345 & 3.68  & 1.268 & 2.5  & Y  \\ 
115272  & Aa & 12 & 0.071 & 21.09 & 1.494 & 38: & N \\
115272  & B &  12 & 0.029 & 3.33  & 0.097 & 0   & N \\  
115272  & C &  4 & 0.390 & 4.09  & 1.597 & 4.1  & Y? \\  
115552  & Aa & 9 & 0.100 & 4.70  & 0.733 & 5.8  & Y \\
115552  & Ab & 9 & 0.120 & 4.73  & 0.568 & 5.9  & Y \\
117596  & Aa & 10 & 0.120 &6.47  & 0.779 & 9.9  & N \\
117596  & Ab & 10 & 0.103 & 6.03 & 0.619 & 9.0  & N 
\enddata 
\end{deluxetable*}

\subsection{Orbit calculation}

As in the  previous papers of this series,  orbital elements and their
errors  were  determined  by   the  least-squares  fits  with  weights
inversely  proportional to  the  adopted errors.   The  IDL code  {\tt
  orbit}\footnote{Codebase:
  \url{http://www.ctio.noao.edu/\~{}atokovin/orbit/}                and
  \url{https://doi.org/10.5281/zenodo.61119} } was used \citep{orbit}.
It  can fit  spectroscopic, visual,  or  combined visual/spectroscopic
orbits. Formal  errors of orbital  elements are determined  from these
fits.    The   elements  of   spectroscopic   orbits   are  given   in
Table~\ref{tab:sborb}, in standard  notation. Its last column contains
the masses  $M \sin^3 i$ for double-lined  binaries.  For single-lined
systems, the  mass of  the primary star  (listed here with  colons) is
estimated from  its absolute $V$  magnitudes, and the minimum  mass of
the secondary that corresponds  to the 90\degr ~inclination is derived
from the orbit.   Table~\ref{tab:rv}, published in full electronically,
provides  individual RVs.  The  HIP number  and the  system identifier
(components  joined by  comma) in  the  first two  columns define  the
pair. Then follow the Julian  date, the RV, its adopted error $\sigma$
(blended CCF dips are assigned  large errors), and the residual to the
orbit (O$-$C).  The  last column specifies to which  component this RV
refers ('a' for the primary and 'b' for the secondary). The 
RVs  of other  visual components  are provided,  for  completeness, in
Table~\ref{tab:rvconst}.  It  contains the HIP  number, the component
letter, the Julian date, and the  RV.

\begin{deluxetable*}{l l cccc ccc c c}    
\tabletypesize{\scriptsize}     
\tablecaption{Spectroscopic orbits
\label{tab:sborb}          }
\tablewidth{0pt}                                   
\tablehead{                                                                     
\colhead{HIP} & 
\colhead{System} & 
\colhead{$P$} & 
\colhead{$T$} & 
\colhead{$e$} & 
\colhead{$\omega_{\rm A}$ } & 
\colhead{$K_1$} & 
\colhead{$K_2$} & 
\colhead{$\gamma$} & 
\colhead{rms$_{1,2}$} &
\colhead{$M_{1,2} \sin^3 i$} 
\\
& & \colhead{(d)} &
\colhead{(+24\,00000)} & &
\colhead{(deg)} & 
\colhead{(km~s$^{-1}$)} &
\colhead{(km~s$^{-1}$)} &
\colhead{(km~s$^{-1}$)} &
\colhead{(km~s$^{-1}$)} &
\colhead{ (${\cal M}_\odot$) } 
}
\startdata
HIP 2863 & Aa,Ab & 4.81156 & 58377.7635 & 0.0 & 0.0 & 55.124               & 88.39 & $-$7.058       & 0.17  & 0.91\\
         &       & $\pm$0.00005 & $\pm$0.0008 & fixed & fixed & $\pm$0.049 & $\pm$0.55 & $\pm$0.036 & 2.23  & 0.56\\
HIP 4974 & Ba,Bb & 14.7104 & 58385.5078 & 0.427 & 114.6 & 53.318          & 70.971 & $-$2.043 & 0.12 & 1.23 \\
         &       & $\pm$0.0002 & $\pm$0.0086 & $\pm$0.001 & $\pm$0.3 & $\pm$0.137 & $\pm$0.194 & $\pm$0.056 & 0.21 & 0.93\\
HIP 8353 & Aa,Ab & 5.3093 & 58371.8164 & 0.000 & 0.0 & 56.721 & \ldots    & 21.043 & 0.83 & 1.6: \\
         &       & $\pm$0.0001 & $\pm$0.0042 & fixed & fixed & $\pm$0.260 & \ldots & $\pm$0.189 & \ldots  & $>$0.86 \\
HIP 28796 &  Aa,Ab &2.5144 & 58308.9570 & 0.038 & 119.9 & 17.140 & \ldots & 21.351 & 0.36 & 1.0: \\
         &          &$\pm$0.0000 & $\pm$0.1716 & $\pm$0.011 & $\pm$24.2 & $\pm$0.370 & \ldots & $\pm$0.161 & \ldots & $>$0.11\\
HIP 35261& Ca,Cb &22.49302 & 58421.602        & 0.2886        & 107.36     & 36.656     &49.285   & $-$6.490  & 0.01 & 0.74 \\
         &       & $\pm$0.00005 & $\pm$0.003  &  $\pm$0.0002   & $\pm$0.05 & $\pm$0.011 & $\pm$0.400 & $\pm$0.006 & 0.73  & 0.56 \\
HIP 92929 & Aa,Ab & 312.279  & 50319.39 & 0.705         & 254.1 & 16.007        & \ldots & 0.163 & 0.05 & 1.1: \\
         &        & $\pm$0.039 & $\pm$0.40 & $\pm$0.007 & $\pm$0.8 & $\pm$0.299 & \ldots & $\pm$0.055 & \ldots & $>$0.50\\
HIP 115272 & Aa,Ab & 3.42691 & 58382.8343 & 0.000 & 0.0 & 50.545            & \ldots & 13.582 & 0.74 & 1.8: \\
         &         & $\pm$0.00001 & $\pm$0.0046 & fixed & fixed & $\pm$0.488 & \ldots & $\pm$0.294 & \ldots & $>$0.65\\
HIP 115552 & Aa,Ab & 17.4779 & 58335.2617 & 0.233 & 8.0 & 60.458 & 62.927 & 16.727 & 0.07 & 1.59 \\
         &          & $\pm$0.0003 & $\pm$0.0184 & $\pm$0.001 & $\pm$0.3 & $\pm$0.068 & $\pm$0.086 & $\pm$0.037 &0.11  & 1.53 \\
HIP 117596 & Aa,Ab & 4.3703 & 58022.1172 & 0.000 & 0.0 & 73.021 & 78.103 & 23.512 & 0.10 & 0.81 \\
         &         & $\pm$0.0000 & $\pm$0.0005 & fixed & fixed & $\pm$0.050 & $\pm$0.050 & $\pm$0.025 & 0.11 & 0.75
\enddata 
\end{deluxetable*}


\begin{deluxetable}{r l c rrr c }    
\tabletypesize{\scriptsize}     
\tablecaption{Radial velocities and residuals (fragment)
\label{tab:rv}          }
\tablewidth{0pt}                                   
\tablehead{                                                                     
\colhead{HIP} & 
\colhead{System} & 
\colhead{Date} & 
\colhead{RV} & 
\colhead{$\sigma$} & 
\colhead{(O$-$C) } &
\colhead{Comp.}  \\
 & & 
\colhead{(JD +2400000)} &
\multicolumn{3}{c}{(km s$^{-1}$)} 
}
\startdata
  2863 &Aa,Ab  & 57985.7850 & $-$60.88 &   0.10 &   0.04 &  a \\ 
  2863 &Aa,Ab  & 57985.7850 &    81.27 &   1.00 &   1.96 &  b \\ 
  2863 &Aa,Ab  & 58341.8440 & $-$61.16 &   0.10 &  $-$0.28 &  a \\
  2863 &Aa,Ab  & 58341.8440 &    81.50 &   1.00 &   2.27 &  b \\
  2863 &Aa,Ab  & 58358.7530 &    45.30 &   0.10 &  $-$0.16 &  a \\
  2863 &Aa,Ab  & 58380.8120 & $-$43.96 &   0.10 &  $-$0.05 &  a \\
  2863 &Aa,Ab  & 58380.8120 &    50.48 &   1.00 &  $-$1.54 &  b 
\enddata 
\end{deluxetable}

\begin{deluxetable}{r l r r }    
\tabletypesize{\scriptsize}     
\tablecaption{Radial velocities of other components
\label{tab:rvconst}          }
\tablewidth{0pt}                                   
\tablehead{                                                                     
\colhead{HIP} & 
\colhead{Comp.} & 
\colhead{Date} & 
\colhead{RV}   \\ 
 & & 
\colhead{(JD $-$2400000)} &
\colhead {(km s$^{-1}$)}  
}
\startdata
2863 & B &   57985.7849 & -3.92 \\
2863 & B &   58341.8436 & -3.94 \\
2863 & B &   58358.7535 & -3.86 \\
2863 & B &   58380.8115 & -3.96 \\
2863 & B &   58382.6616 & -3.92 \\
2863 & B &   58384.6394 & -3.98 \\
2863 & B &   58390.7608 & -3.79 \\
2863 & B &   58393.6771 & -3.75 \\
2863 & B &   58397.6459 & -3.94 \\
4974 & A &   57985.7901 & -2.92 \\ 
4974 & A &   57986.8340 & -2.95 \\ 
4974 & A &   58358.7681 & -2.92 \\ 
4974 & A &   58370.8620 & -2.94 \\ 
4974 & A &   58380.8179 & -3.03 \\ 
4974 & A &   58382.6662 & -2.84 \\ 
4974 & A &   58383.8077 & -2.75 \\ 
4974 & A &   58384.6054 & -2.90 \\ 
4974 & A &   58385.8176 & -2.88 \\ 
4974 & A &   58393.6137 & -2.90 \\ 
8353 & B &   57985.7997 & 21.95 \\ 
8353 & B &   57986.8466 & 22.36 \\ 
8353 & B &   58370.8769 & 22.66 \\ 
8353 & B &   58380.8455 & 21.54 \\ 
8353 & B &   58382.8260 & 22.28 \\ 
8353 & B &   58384.6274 & 22.27 \\ 
8353 & B &   58385.6427 & 22.34 \\ 
8353 & B &   58393.7654 & 22.22 \\ 
28796& B &   57266.8736 & 14.39 \\ 
28796& B &   57319.6842 & 14.43 \\ 
28796& B &   57364.6872 & 14.43 \\ 
28796& B &   57374.6120 & 14.45 \\ 
28796& B &   57985.8975 & 14.44 \\ 
28796& B &   57986.9151 & 14.48 \\ 
28796& B &   58193.4814 & 14.29 \\ 
28796& B &   58195.4857 & 14.53 \\ 
28796& B &   58382.8844 & 14.48 \\ 
28796& B &   58395.8906 & 14.53 \\  
28796& B &   58407.8409 & 14.13 \\ 
28796& B &   58408.7916 & 14.47 \\  
28796& B &   58409.8807 & 14.50 \\  
115272& B &   57984.7192 & 11.89  \\ 
115272& B &   57986.6695 & 11.42  \\  
115272& B &   58340.7511 & 12.56   \\ 
115272& B &   58341.8367 & 12.33  \\ 
115272& B &   58342.7908 & 12.59  \\ 
115272& B &   58356.6185 & 11.15   \\ 
115272& B &   58380.7650 & 11.23  \\ 
115272& B &   58382.6583 & 11.77   \\ 
115272& B &   58383.6587 & 12.08   \\ 
115272& B &   58384.6362 & 12.14  \\  
115272& B &   58385.6255 & 11.46  \\  
115272& B &   58390.6450 & 11.48  
\enddata 
\end{deluxetable}

\section{Individual objects}
\label{sec:obj}

For  each observed system,  the corresponding  Figure shows  a typical
 CCF (the Julian date and individual components are marked
on the plot) together with the RV curve representing the orbit. In the
RV curves, squares denote  the primary component, triangles denote the
secondary component, while the full and dashed lines plot the
orbit. Masses of stars are estimated from absolute magnitudes, orbital
periods of wide pairs from their projected separations
\citep[see][]{MSC}.  

\subsection{HIP 2863 (Triple)}

\begin{figure}
\epsscale{1.0}
\plotone{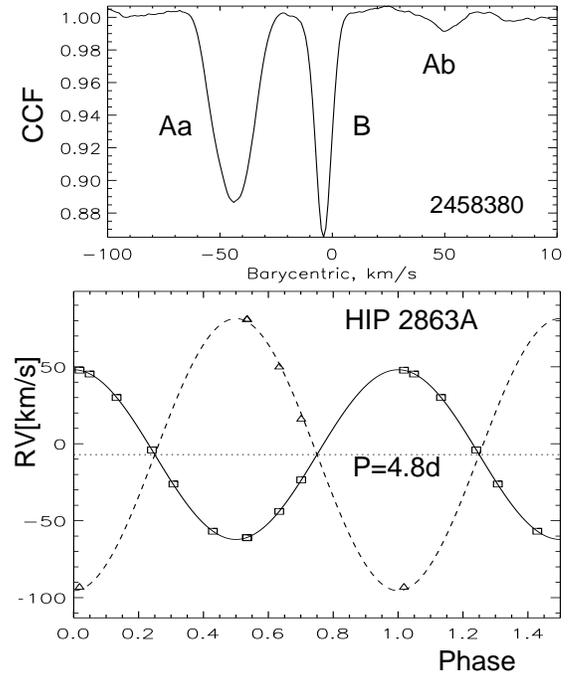}
\caption{CCF (top)  and RV curve (bottom)  of HIP 2863  Aa,Ab. In this
  and the following Figures, the upper panel shows a typical CCF, with
  the  date of  observation indicated.  The lower  panel  presents the
  phased RV  curve where the full  line and squares  denote the orbit
  and  RVs of  the primary  component, the  dashed line  and triangles
  refer to the secondary component.
\label{fig:2863}
}
\end{figure}

The 0\farcs4  visual binary  RST~5183 with fast proper motion  (PM) was  discovered by
R.A.~Rossiter  in 1945.  Since then,  it moved by  12\degr ~in  angle and
opened up to  0\farcs7 separation. The orbital period  of this pair is
estimated at $\sim$400 years.

\citet{N04} found a large  RV variation and double lines. Observations
with CHIRON conducted since 2017  show a superposition of two spectra,
one with narrow  lines and a constant RV of  $-$3.85 \kms (rms scatter
0.08 \kms),  another with  a fast RV  variation and wider  lines.  The
latter belongs  to the bright component  A of the  visual binary.  Its
orbit with  a period of 4.8 days  is circular (Figure~\ref{fig:2863}).
The center-of-mass  velocity, $-7.07$ \kms,  differs from the RV  of B
owing to the motion in the visual orbit.

Some (but not  all) CCFs contain a very weak  dip corresponding to Ab,
originally  ignored. Fitting  this dip  defines the  RV
amplitude  of Ab  and   the  mass ratio  $q_{\rm Aa,Ab}  =
0.62$. The mass of Aa deduced from its absolute $V$ magnitude is about
1.3 \msun,  hence the  mass of  Ab is 0.82  \msun. Comparison  with $M
\sin^3 i$ leads  to the orbital inclination $i_{\rm  Aa,Ab} = 62^\circ$.
The mass  of B is intermediate,  0.95 \msun. Slow axial  rotation of B
and the  absence of lithium  in the spectrum  indicate an old  age, in
harmony  with the fast PM and  the metallicity  of
[Fe/H]=$-$0.37 \citep{N04}.

\subsection{HIP 4974 (Triple)}

\begin{figure}
\epsscale{1.0}
\plotone{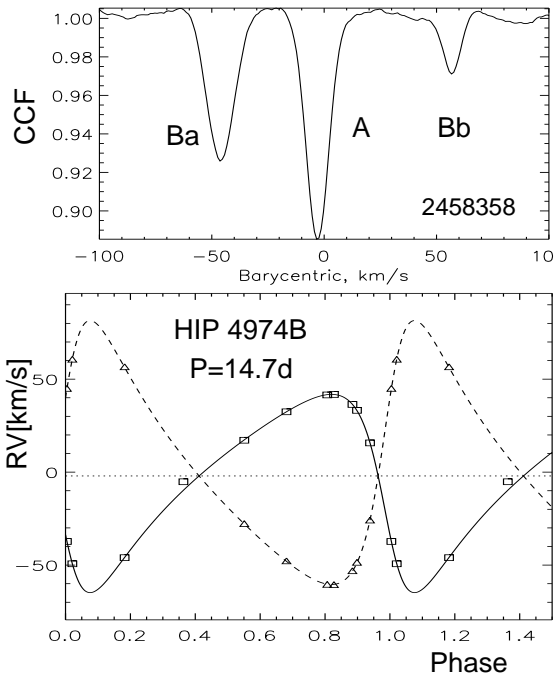}
\caption{CCF (top)  and RV curve (bottom) of HIP 4974.
\label{fig:4974}
}
\end{figure}

Like  the previous  object, this  is a  classical visual  binary B~649,
known since  1927. The orbital  separation of 0\farcs8  corresponds to
the period  of $\sim$250 years;  the stars are of  similar brightness,
$\Delta V_{\rm AB}  = 0.40$ mag. Only a small arc  of the outer orbit
is observed. The {\it Gaia} DR2  parallaxes of A and B are discordant,
5.14$\pm$0.46 and 16.09$\pm$1.09  mas respectively. The large parallax
errors indicate a problem in  the data reduction, presumably caused by
the small separation.  The {\it  Hipparcos} parallax of AB, 13.87 mas,
is adopted here. However, it places the unresolved star below the main
sequence, so the actual parallax is likely smaller. 

The large  RV variation reported by  \citet{N04} prompted observations
of this system  with CHIRON. Independently, eclipses with  a period of
14.7104   days  and   a  depth   of  0.22   mag  were   discovered  by
\citet{Otero2004}.  The variable star  designation is SI~Scl.

The  spectrum is  triple-lined (Figure~\ref{fig:4974}).   The stronger
lines, attributed here to the  the visual component A, are stationary,
with an RV of $-$2.9 \kms.  They are accompanied by the rapidly moving
lines of  the subsystem  Ba,Bb.  Its spectroscopic  period, determined
independently,  matches perfectly the  photometric period.   The orbit
has an  eccentricity of  0.43, while the  mass ratio $q_{\rm  Ba,Bb} =
0.75$. Considering  the uncertain  distance, estimates of  masses from
luminosity are questionable. However, the presence of eclipses implies
that the spectroscopic  masses of Ba and Bb, 1.23  and 0.93 \msun, are
close to the  actual masses.  A joint analysis of  the light curve and
RVs will lead to the  accurate mass measurement for the eclipsing pair.
The presence of lithium in the spectrum of component A and the spatial
velocity $(U,V,W)  = (-3.4,  -17.2, -10.7)$ \kms  (the $U$ axis  is directed
away from the  Galactic center) indicate that this  multiple system is
relatively young.

It  is not  certain that  the inner  subsystem belongs  to  the visual
secondary B. The sum of the CCF  areas of Ba and Bb is larger than the
area of A  by 9\%. However, for the spectral type  around F5V the CCF
area  depends on the  effective temperature,  being larger  for cooler
stars.  Direct differential  photometry of A and B  during eclipses or
displacement of  the photo-center of  the unresolved images of  AB can
confirm the attribution of the spectroscopic pair to the component B.

\subsection{HIP 8353 (Triple)}

\begin{figure}
\epsscale{1.0}
\plotone{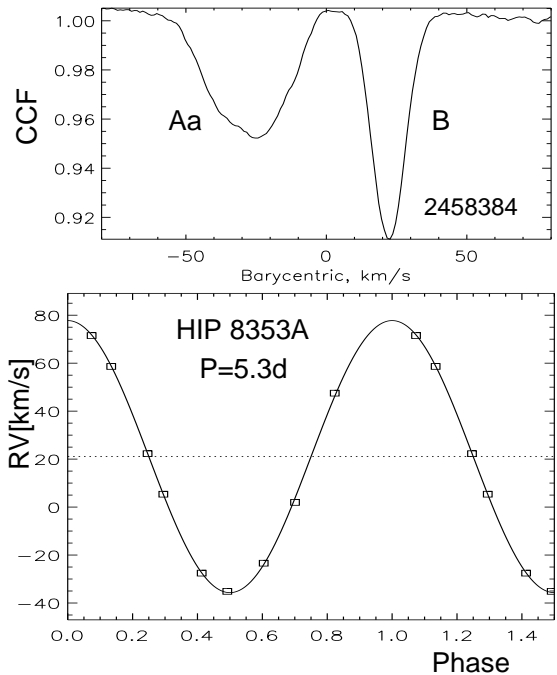}
\caption{CCF (top)  and RV curve (bottom) of HIP 8353.
\label{fig:8353}
}
\end{figure}

This  object is a  close visual  binary I~52  for which  a preliminary
orbit with $P=224$ years, a  semimajor axis of 0\farcs372, and a large
eccentricity  of  0.847 has  been  computed  by \citet{Doc2016a}.  The
components A and  B are similar, $\Delta V_{\rm  AB} \approx 0.5$ mag.
Owing to  the resolved nature of  the source, {\it Gaia}  does not yet
provide  the  parallax, and  I  adopt  the  original {\it  Hipparcos}
parallax  of 5.3$\pm$1.2  mas. The  estimated mass  sum and  the  visual orbit
correspond to the dynamical parallax of 6.4 mas.

As noted  by \citet{N04}, the spectrum is  double-lined.  I attribute
the narrow  stationary lines to the  visual component B  and the wider
moving  lines  to  the  component  Aa, although  this  choice  is  not
absolutely certain.  The  mean RV of B is 22.15  \kms with rms scatter
of 0.28 \kms.   The circular orbit of Aa,Ab with a  period of 5.3 days
is  computed here  (Figure~\ref{fig:8353}).  Large  residuals  of 0.83
\kms  are  explained by  the  wide and  shallow  dip  of Aa,  probably
distorted by star-spots.   If the masses of Aa and B  are 1.67 and 1.50
\msun, as deduced from their  absolute magnitudes, the minimum mass of
Ab is 0.81 \msun.  As its lines are not detectable, it could be either
a low-mass dwarf or a cold white dwarf. The Vizier photometry tool\footnote{ 
See \url{http://vizier.u-strasbg.fr/vizier/sed/}} does not indicate
any ultra-violet excess. 


\subsection{HIP 28796 (Triple)}

\begin{figure}
\epsscale{1.0}
\plotone{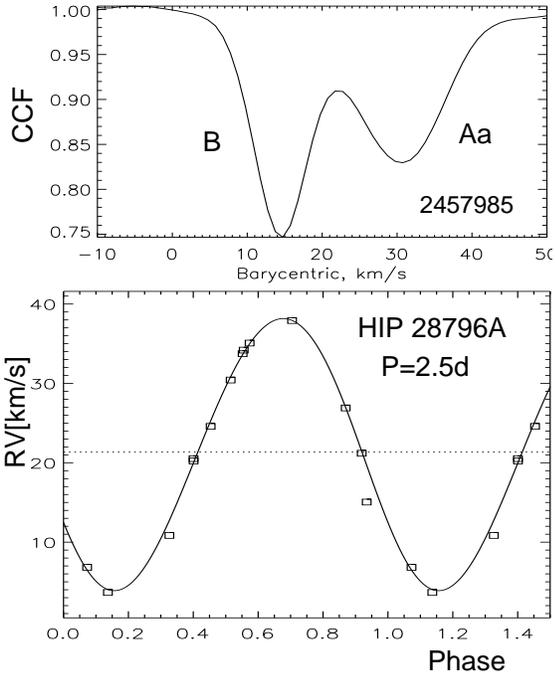}
\caption{CCF (top)  and RV curve (bottom) of HIP 28796.
\label{fig:28796}
}
\end{figure}

This is  a nearby (distance  30.7 pc) chromospherically  active triple
system V575~Pup  (HR 2162).  Activity  and X-ray flux suggest  a young
age,  and  membership in  the  Castor  moving  group was  proposed  by
\citet{Caballero2010}.   However, the  spatial  velocity of  HIP~28796
$(U,V,W) = (-2.5, -11.7, -23.2)$ is quite distinct from the motion of
that group.  \citet{Plavchan2009} searched for excess thermal emission
from a debris disk but have not found any.

The visual  orbit of the outer  pair A,B is uncertain;  the latest one
\citep{Tok2014a} has the period of 916 years and the semimajor axis of
4\farcs578.  The large RV  variability detected by \citet{N04} implied
the existence of a  spectroscopic subsystem. Its orbit determined here
has  a period of  2.51 days  (Figure~\ref{fig:28796}).  Interestingly,
the eccentricity  is marginally significant  ($e = 0.038  \pm 0.011$),
despite  the  short  period.    The  visual  binary  has  currently  a
separation  of  2\farcs6,  allowing  to  differentiate  the  light  of
components in  the CHIRON fiber  under good seeing.   Several resolved
spectra of both  components were taken; in the  remaining spectra, the
light of A  and B is mixed and  the CCF has a double dip,  as shown in
Figure~\ref{fig:28796}.

Synchronous rotation of the star  Aa, of one solar radius, corresponds
to  the equatorial  speed of  20 \kms.  Comparison with  the projected
rotation of  8.8 \kms  provides an  estimate of the  orbital inclination
$i_{\rm  Aa,Ab} \approx  26\degr$.  The  minimum  mass of  Ab is  0.11
\msun; accounting for the inclination,  the true  mass  is about  0.3  \msun.

Slow axial rotation of the component  B and the absence of the lithium
line  in the  spectra suggest  that this  triple system  is relatively
old.  The fast  rotation  of  Aa and  its  chromospheric activity  are
sustained  by the  angular  momentum of  the  close binary.   

The RV  of B is constant  at 14.45 \kms  with the rms scatter  of 0.11
\kms, while the  center-of-mass velocity of Aa,Ab is  21.35 \kms.  The
difference RV(B)$-$RV(A)=$-$6.9$\pm$0.2 \kms ~might seem too large for
the millennium-long outer period.   The orbit was re-fitted to evaluate
its  reliability, resulting  in $P_{\rm  A,B}  = 947  \pm 289$  years,
$a_{\rm A,B}  = 4\farcs48 \pm 0\farcs76$,  and $e_{\rm A,B}  = 0.43 \pm
0.11$. The {\it Gaia} parallaxes of  A and B are accurate and mutually
consistent,  $\overline{\omega}_{\rm  A} =  32.51  \pm  0.03$ mas  and
$\overline{\omega}_{\rm  B}  =  32.47  \pm  0.03$  mas.   The  average
parallax and  the still uncertain orbit lead to  the mass sum  of 3.11$\pm$0.19 \msun,
larger than  the estimated mass  sum of 2.25  \msun.  The pair  A,B is
presently  located near  the node  of its  orbit and  the  computed RV
difference between  B and  A is $-$6.1  \kms, only slightly  less than
actually  measured.   The  true  ascending   node  of  the   orbit  is
$\Omega_{\rm A}  = 122\degr$.   The RV of  B is constant  during three
years  covered by  CHIRON.  The  observed motion  of the  pair  A,B is
smooth, without any hint  of astrometric perturbation.  Therefore, the
existence of additional companions  to B is highly unlikely. 


\subsection{HIP 35261 (Quadruple)}

\begin{figure}
\epsscale{1.0}
\plotone{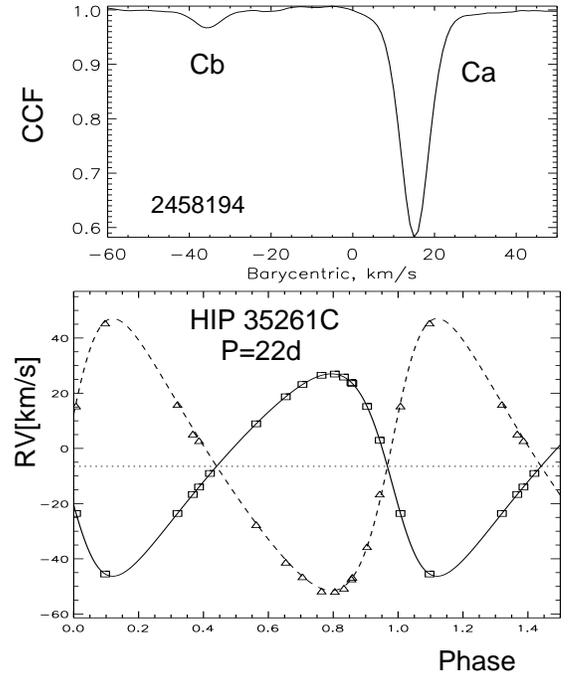}
\caption{CCF (top)  and RV curve (bottom) of HIP 35261C.
\label{fig:35261}
}
\end{figure}

This is a  nearby (53 pc) quadruple system.   The outer 15\farcs9 pair
STF~1064 AB,C  is known  since 1831 and  is definitely  physical.  Its
main component is  the visual binary A~2123 A,B with  a known orbit of
80 year period and the semimajor  axis of 0\farcs64.  This pair is not
present in the {\it Gaia} DR2, while its HIP2 parallax is 21.6$\pm$2.1
mas.   On the  other hand,  the component  C has  a  well-measured DR2
parallax of 19.03$\pm$0.07 mas.  \citet{survey} discovered that C is a
double-lined  binary with  very unequal  dips.  Here  I  compute its
spectroscopic  orbit with  $P=22$ days  (Figure~\ref{fig:35261}) using
RVs measured in  2015--2018.  The residuals  for the component
Ca are only 0.013 \kms, owing to its narrow and deep spectral lines. 

Comparison  between spectroscopic  and  estimated masses  of the  pair
Ca,Cb leads  to an orbital inclination of  74\degr. Interestingly, the
inclination  of A,B  is  78\degr; the  two  orbits could  be (but  not
necessarily are) coplanar. The component Ca is, apparently, a normal
slowly rotating K1V dwarf. The dip of Cb is so weak that its
parameters are poorly measured.

\subsection{HIP 92929 (Triple)}

\begin{figure}
\epsscale{1.0}
\plotone{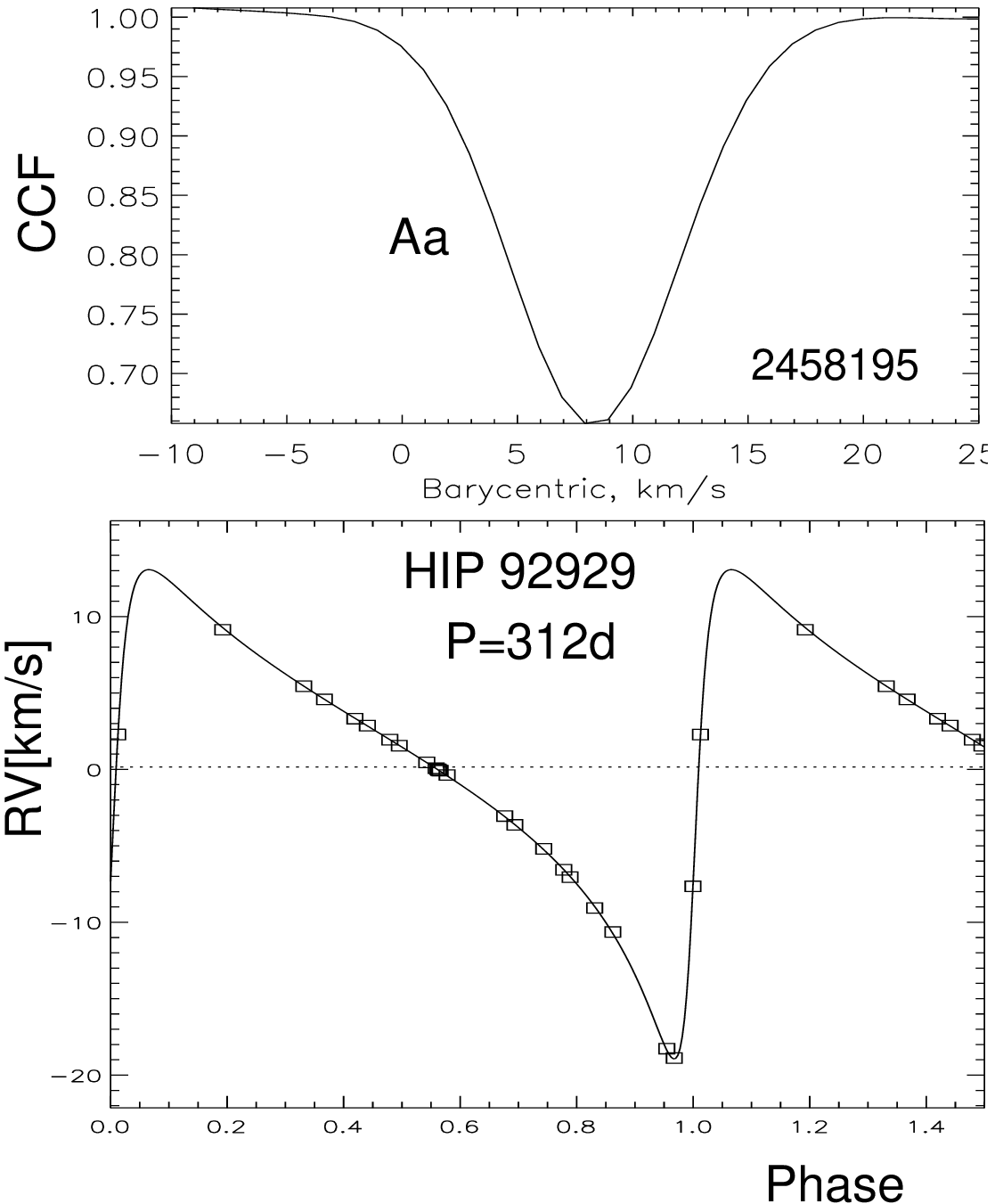}
\caption{CCF (top)  and RV curve (bottom) of HIP 92929.
\label{fig:92929}
}
\end{figure}

The  outer system  is  a 6\farcs2  visual  binary B~413  with a  large
$\Delta V  = 5.6$  mag, an  estimated period of  $\sim$3 kyr,  and the
correspondingly slow relative motion. The RV variation of the brighter
component  A  was  noted  by  \citet{Jenkins2015}  during  search  for
exo-planets.  They determined a preliminary  orbit with $P = 312$ days
and $e=0.75$,  although the coverage  of the RV curve  was incomplete.
The period  is confirmed by CHIRON observations,  but the eccentricity
is different, $e=0.70$; it  is well constrained by RVs on
the rising branch of  the RV curve in Figure~\ref{fig:92929}.  Precise
RVs from  \citet{Jenkins2015} match the  CHIRON RVs very  well because
the orbit computed from the  combined data has residuals of only 0.045
\kms, without any adjustment of  the velocity zero point.  The minimum
mass of the secondary component Ab is 0.5 \msun.

The narrow CCF dip of Aa has  a large depth and corresponds to $V \sin
i  =  2.5$ \kms.   Despite  the  slow  projected axial  rotation,  the
spectrum contains a strong lithium line with an equivalent width of 38
m\AA. The  spatial velocity  $(U,V,W) = (-2.7,  11.9, -2.7)$ \kms  does not
match any known kinematic group.



\subsection{HIP 115272 (Quadruple)}

\begin{figure}
\epsscale{1.0}
\plotone{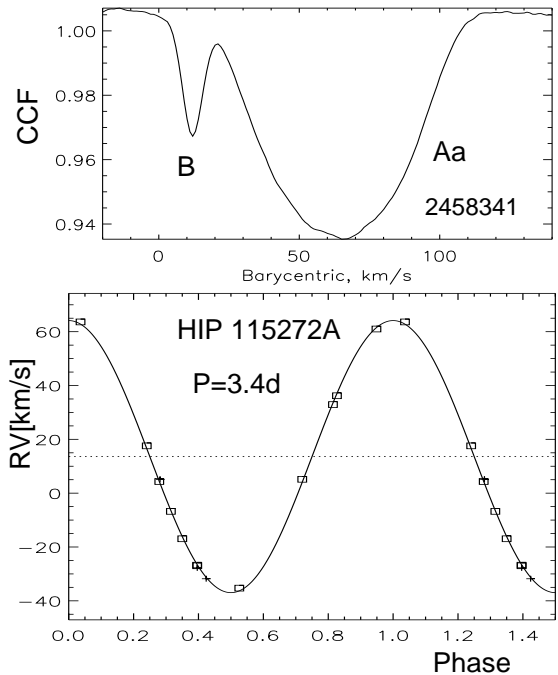}
\caption{CCF (top)  and RV curve (bottom) of HIP 115272. The three RVs
  measured by \citet{N85} are plotted as crosses.
\label{fig:115272}
}
\end{figure}

This bright  quadruple system is located  at 75\,pc from  the Sun. The
outer  17\arcsec ~pair AB,C  (DUN~248) is  composed of  HIP~115272 and
HIP~115269, with common  PMs, parallaxes, and RVs. The  period of AB,C
is on  the order of 20  kyr. The mean  RV of the component  C measured
with CHIRON is 14.301 \kms with the rms scatter of 0.016 \kms.  At the
next  hierarchical  level  we  find  the 1\farcs3  visual  binary  A,B
(RST~5560) with $\Delta  V = 2$ mag, an  estimated period of $\sim$500
years, and  unknown orbit.  Double lines  in the spectrum  of this star
were noted by \citet{N85}, suggesting  that it contains an even closer
spectroscopic binary.

The CHIRON spectra of AB show double lines. The set of weaker lines is
stationary with  the mean RV of  11.94 \kms (rms scatter  0.47 \kms); it
corresponds to the visual secondary component B. The stronger lines of
Aa   move  with   the  period   of  3.4   days  on   a   circular  orbit
(Figure~\ref{fig:115272}).  The  three RVs from  \citet{N85} match the
orbit  and  improve the  accuracy  of the  period.   The  mass of  the
metallic-line  F5m component  Aa, estimated  crudely from  its absolute
magnitude,  is  1.8  \msun,  hence  the  minimum mass  of  Ab  is  0.65
\msun. The  component C is located  on the main sequence,  while AB is
above  it.  With the  Aa radius  of 2.75  $R_\odot$ estimated  by {\it
  Gaia}, the  synchronous equatorial velocity  is 40 \kms, in  rough
agreement with the actual width of the CCF dip.

\subsection{HIP 115552 (Triple) }

\begin{figure}
\epsscale{1.0}
\plotone{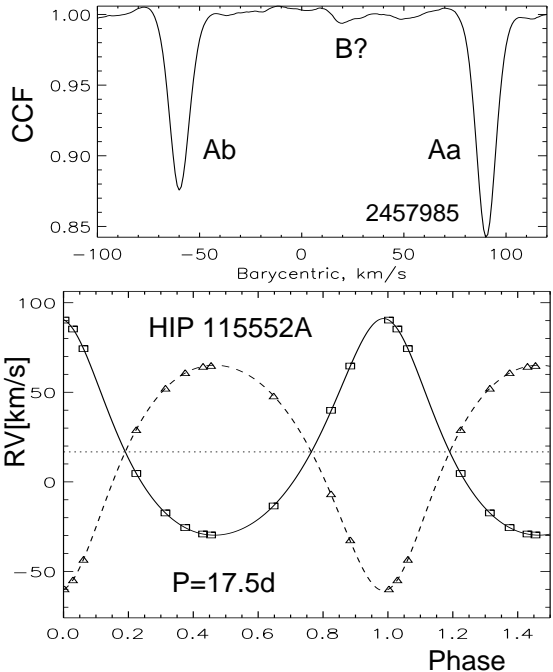}
\caption{CCF (top)  and RV curve (bottom) of HIP 115552.
\label{fig:115552}
}
\end{figure}

This triple system consists of the 1\farcs5 visual binary A,B (BU~719,
estimated period 1.5 kyr) and the double-lined spectroscopic subsystem
Aa,Ab  with the period  of 17.5  days and  a moderately  eccentric orbit
(Figure~\ref{fig:115552}).  The spectroscopic subsystem was discovered
by \citet{N04}. The visual secondary B is 3 mag fainter than A and its
lines are  not detected in  the spectra, although the  persistent weak
detail  of  the CCFs  near  the  center-of-mass  velocity may  actually
correspond to B.

The two  stars Aa and  Ab, of F8V  spectral type, are nearly  equal, the
mass ratio is $q_{\rm Aa,Ab}  = 0.96$. Their estimated masses are 1.65
\msun  each, close  to the  spectroscopic  masses $M  \sin^3i$ of  1.6
\msun.  Therefore, the  orbital inclination  is large,  $i_{\rm Aa,Ab}
\approx 80^\circ$. 

Stellar rotation  in a binary with  an orbital period of  17.5 days is
not expected to be synchronous. Indeed,  both Aa and Ab rotate with $V
\sin i  = 5.8$ \kms, faster  than the pseudo-synchronous  speed of 3.8
\kms, calculated for  $ 1 R_\odot$ radius. The  presence of lithium in
the spectra  of both components  points to the  young age, as  well as
their location above  the main sequence. These stars  are too luminous
for  their spectral  type F8V  (which matches  the $V-K$  color)  by a
factor of  $\sim$4.  However, the  spatial velocity $(U,V,W)  = (-5.5,
-0.1, -24.0)$ \kms does not match any known group of young stars.


\subsection{HIP 117596 (Triple) }

\begin{figure}
\epsscale{1.0}
\plotone{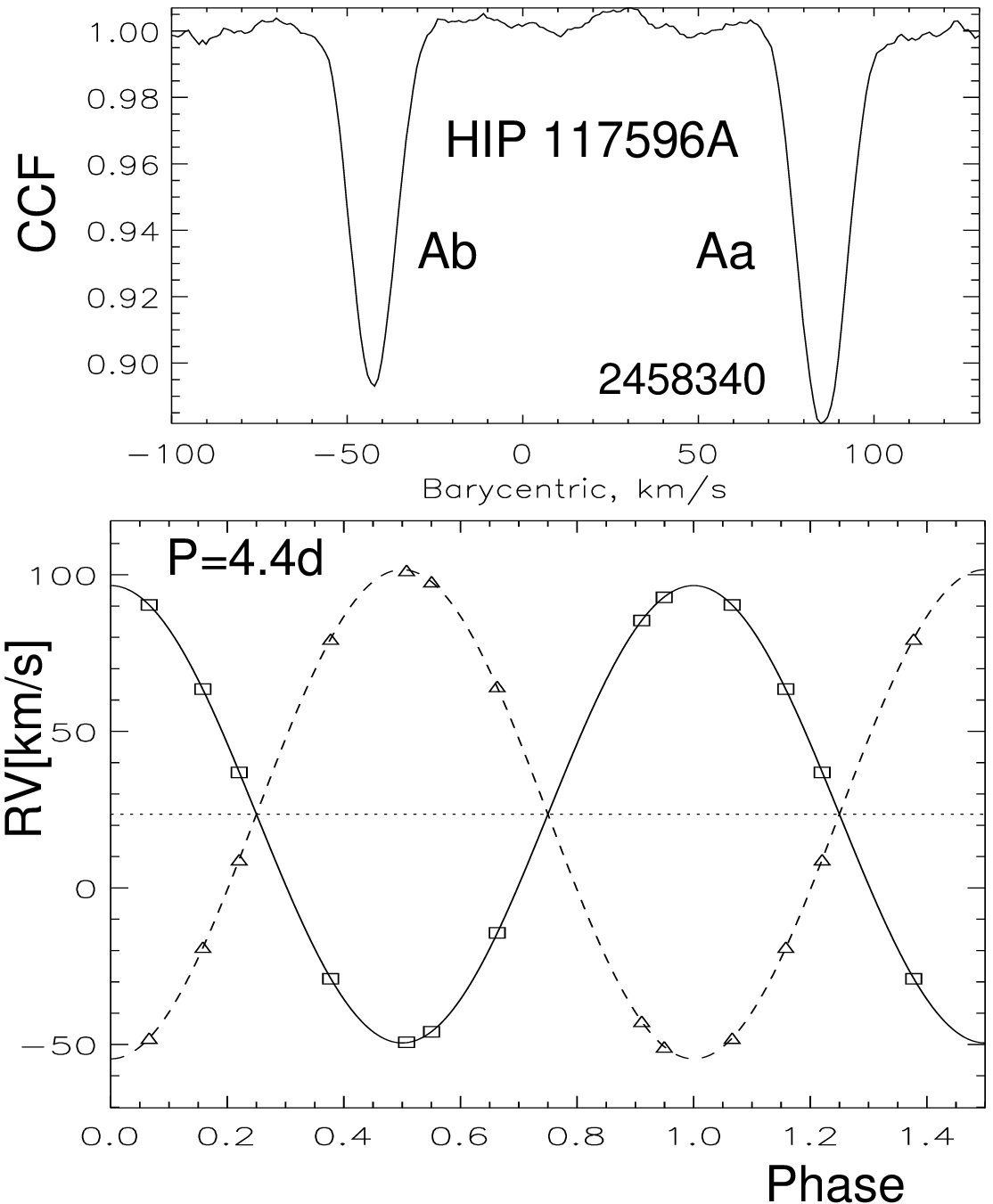}
\caption{CCF (top)  and RV curve (bottom) of HIP 1117596.
\label{fig:117596}
}
\end{figure}

The outer pair A,B is  wide (30\farcs7, UC~5034). The faint ($V=15.55$
mag) star B has common PM  and parallax with A according to {\it Gaia}
DR2. Its color and luminosity match a dwarf of spectral type M3V.  The
estimated period  of A,B  is of  the order of  50\,kyr. The  system is
moderately metal-poor,  [Fe/H] =  $-$0.37 \citep{N04}. The  absence of
lithium indicates that it is not young.

Double lines in the spectrum of A have been noted by \citet{N04}.  The
system is featured in the catalog of chromospherically active binaries
by  \citet{Eker2008}, although they do not  provide the  period.  The
circular  orbit  of  Aa,Ab  with  $P=4.37$  days  is  determined  here
(Figure~\ref{fig:117596}). This  is a twin pair with  $q_{\rm Aa,Ab} =
0.93$. The  estimated masses of 0.99  and 0.93 solar are  close to the
spectroscopic masses,  hence the inclination is  large, $i_{\rm Aa,Ab}
\approx  70^\circ$. The  synchronous equatorial  rotation  of $\sim$11
\kms roughly matches the measured $V \sin i$.


\section{Summary}
\label{sec:sum}

Monitoring  of  hierarchical  systems  with  CHIRON  has  led  to  the
determination  of 23  orbits published  in the papers 1--4;  the paper  5
brings the total to 32. The cumulative contribution of this program to
the data on nearby low-mass  hierarchies is far from  negligible.
For  example,  the number  of  spectroscopic  subsystems with  unknown
periods in the 67-pc sample  of solar-type stars \citep{FG67a} was 61.
After  excluding long-period  (e.g.  astrometric)  subsystems  and the
recently determined orbits, the number of unknown short periods before
this publication was 13, and four of those are defined here.  Complete
coverage of the short-period hierarchies in this volume-limited sample
can be reached soon.  It  will establish the unbiased distributions of
periods  and mass ratios  of the  inner subsystems  that will  help to
clarify the still debated origin  of close binaries. The preference of
close binaries to be members of  hierarchies is a known fact which may
be explained in several different ways \citep{Moe2018}.

In  this  survey-type  work  I inevitably  encounter  interesting  and
unusual systems,  such as the  young quadruple system HD~86588  with a
short period and, yet, eccentric orbit \citep{young}.  Several compact
triple systems where both outer  and inner spectroscopic orbits can be
derived  will be presented  in the  future paper  of this  series when
their outer  orbits are covered.  Unusual  hierarchies give additional
clues to the origin of stellar systems.

\acknowledgements

I thank the operator of  the 1.5-m telescope R.~Hinohosa for executing
observations  of  this  program  and  L.~Paredes  for  scheduling  and
pipeline processing.  Re-opening of CHIRON  in 2017 was largely due to
the enthusiasm and energy of T.~Henry.

This work  used the  SIMBAD service operated  by Centre  des Donn\'ees
Stellaires  (Strasbourg, France),  bibliographic  references from  the
Astrophysics Data  System maintained  by SAO/NASA, and  the Washington
Double Star Catalog maintained at USNO.
This work has made use of data from the European Space Agency (ESA) mission
{\it Gaia} (\url{https://www.cosmos.esa.int/gaia}), processed by the {\it Gaia}
Data Processing and Analysis Consortium (DPAC,
\url{https://www.cosmos.esa.int/web/gaia/dpac/consortium}). Funding for the DPAC
has been provided by national institutions, in particular the institutions
participating in the {\it Gaia} Multilateral Agreement.

{\it Facilities:}  \facility{CTIO:1.5m}










\end{document}